\documentstyle[12pt]{article}

\newcommand{\beq}{\begin{equation}}
\newcommand{\eeq}{\end{equation}}

\begin{document}

\begin{titlepage}

\thispagestyle{empty}

\begin{flushright}
BI-TP 94/27
\end{flushright}
\begin{flushright}
May 1994
\end{flushright}

\vspace{0.5truecm}

\begin{center}
\large\bf{Effective Scalar Field Theory\\
for the Electroweak Phase Transition\\
}
\end{center}

\vspace{0.5truecm}

\begin{center}
\large{F. Karsch$^{1}$, T. Neuhaus$^{1}$ and A. Patk\'os$^{2}$
}
\end{center}

\footnotetext[1]{{\em
 Fakult\"at f\"ur Physik, Universit\"at Bielefeld,
 D-33615 Bielefeld, FRG}}
\footnotetext[2]{{\em
 Department of Atomic Physics, E\"otv\"os University,
 H-1088, Puskin u. 5-7, Budapest, Hungary}}

\vspace{0.5truecm}

\begin{center}
\large{Abstract:}
\end{center}

    We investigate an effective model
for the finite temperature symmetry restoration
phase transition of the electroweak theory. It is
obtained by dimensional reduction
of the $3+1$ dimensional full theory and by subsequent integration
over all static  gauge degrees of freedom. The resulting theory
corresponds to a $3$-dimensional $O(4)$ ferromagnet
containing cubic and quartic terms of the field in its
potential function. Possible nonperturbative effects of a magnetic
screening mass are parametrically included in the potential.
We analyse the theory using mean field
and numerical Monte Carlo (MC) simulation methods. At the
value of the physical Higgs mass, $m_H=37~{\rm GeV}$, considered
in the present investigation, we find
a discontinuous symmetry restoring phase transition. We determine
the critical temperature, order parameter jump, interface tension
and latent
heat characteristics of the transition. The Monte Carlo results indicate
a somewhat weaker first order phase transition as compared to the
mean field treatment, demonstrating
that non-perturbative fluctuations of the
Higgs field are relevant. This effect is especially important for the
interface tension.
Any observation of hard first order transition could result only from
non-perturbative effects related to the gauge degrees of freedom.

\end{titlepage}

\newpage

\section{Introduction}

The tiny matter-antimatter asymmetry observed on cosmic scales requires
explanation in the framework of standard cosmology.
Baryon number violating processes are known to occur in the electro-weak
theory and are, in fact, expected to occur frequently at
high temperatures in equilibrium. The occurence of a non-vanishing
matter-antimatter
asymmetry, however, is expected to be possible only, if the universe
evolved through some non-equilibrium stages.
If the electroweak phase transition
was of first order (discontinuous) nature, then
at temperatures $T_c \sim 100~{\rm GeV}$ there might have been a chance for
developing the observed baryon asymmetry of the universe.
The very active present day investigation of this question is based
predominantly on
a perturbative evaluation of the effective potential of the Higgs-field.

The interest in non-perturbative studies stems from two sources - the
generation of a magnetic mass as well as the breakdown of the perturbative
treatment in the case of a weakly first order phase transition.
A non-perturbatively generated screening mass of the magnetic part of
the gauge field fluctuations is expected to weaken the discontinuous
nature of the transition. Eventually
this could lead to a second order transition, if
the magnetic screening mass becomes too large \cite{Dine}.
Thus the perturbative calculations \cite{bucha,espinosa}
lead to constraints on the maximal magnetic screening
mass allowed for a first order transition and also lead to other predictions,
like for instance the latent heat and surface tension, important
for discussing the kinetics of the phase transformation.

The validity of any perturbative treatment of the transition has been
questioned for first order transitions where near $T_c$ the ratio
$T/m(\Phi_{\rm min})$ becomes large (weakly first order transition). In
this case the fluctuations of the light modes (especially of the
Goldstone modes) might require special considerations. Very recently
various exploratory studies have been performed, in which it has been
attempted to
apply renormalization group theory of critical phenomena to this
situation \cite{gleiser,alford,arnold}. However, one has to emphasize
that techniques like the $\epsilon$-expansion were successful in finding
universal characterisations, mainly critical exponents. Their success
for calculating critical temperatures, order parameter discontinuities
or surface tensions is by no means guaranteed.

Under these circumstances numerical simulations seem to be an especially
valuable source of information. Exploratory investigations of the full,
4-d finite temperature theory met difficulties because of the large
fluctuations due to the weakly coupled nature of the phenomena
\cite{bunk}. In this situation dimensional reduction might prove to be
of important practical help \cite{kajantie,farakos}.

In this work
we study the influence of the finite temperature fluctuations of the
Higgs-field on these parameters. In particular we will examine an effective
3-d theory for the Higgs-field, which has been obtained from the
(3+1)-dimensional SU(2)-Higgs model in two steps:

\begin{description}
\item{i)}{ ~~Dimensional reduction: One
integrates over all non-static Matsubara
fields at one-loop level. Since these modes are massive, no
infrared sensitivity
is expected, their perturbative integration seems to be well-founded.}
\item{ii)}{ ~Elimination of gauge degrees of freedom: The
theory resulting from
the first step
is a 3-dimensional Gauge-Higgs model, where in addition, also an isovector
field, the fourth component of the gauge fields, is present. In order to reduce
the theory further one integrates over the magnetic gauge degrees of freedom
and the isovector scalar. According to the improved perturbative
treatments, the infrared stability can be ensured, if one includes into
the result of the "naive" 3-dimensional 1-loop calculation a magnetic and
an electric screening mass. Especially the first of them lacks, however, firm
theoretical basis.}
\end{description}

We remark that the resulting effective scalar theory has been used
recently also for the investigation of bubble nucleation in the
electroweak phase transition \cite{buchb}.

In section 2 we first present the result of step i) for an SU(2) Higgs-model
with N doublets:
\begin{equation}
S=\int_{0}^{\beta}d\tau \int d^{3}x\bigl [{1\over 4}F_{\mu \nu}^{a}
F_{\mu \nu}^{a}+
{1\over 2}(D_{\mu}\Phi_q)^{+}(D_{\mu}\Phi_q)+{1\over 2}m^{2}\Phi_q^{+}\Phi_q
+{\lambda\over 24}(\Phi_q^{+}\Phi_q)^{2}\bigr ]
+{\rm c.t.},
\end{equation}
$(q=1,...,N; \mu =1,..,4; a=1,2,3; D_{\mu}\Phi =
(\partial_{\mu}+igA_{\mu}^{a}\tau^{a}/2)\Phi)$.

 In the limiting case N=$\infty$ step ii) can be
performed exactly and a pure scalar action of very similar form to
that actually studied in later sections is found. Also, some evidence will
be presented from the perturbative analysis of the 3-d effective high
temperature action of the N=1 Higgs model that in the coupling region
corresponding to small Higgs mass values the pure scalar model might
offer considerable insight into the physics of the electroweak phase
transition.

The discretisation of the effective model is described in section 3.
The continuum limit of its mean field solution will be discussed in
order to clarify the strategy of the non-perturbative investigations.
Special attention will be paid to the uncertainties in taking the
continuum limit, arising from the application of different renormalisation
conditions.

Section 4 presents a detailed discussion of the Monte Carlo
simulation of the effective scalar model. Several physical quantities
relevant to the phase transition will be stochastically evaluated and
compared carefully to the information available from the literature. In
the simulation the mass of the Higgs field was chosen approximately
$m_H(T=0)\sim 35~{\rm GeV}$.
Conclusions will be drawn in section 5.
\bigskip

\section{Derivation of the Effective Scalar Action}

In this section we outline the  derivation of the effective 3-d action for
the SU(2) Higgs model with N scalar doublets  on the 1-loop level.
It represents a natural framework
for the investigation of the finite temperature electroweak phase transition
in an appropriate {\it large N} limit \cite{arnold,jain}.
It turns out that the resulting 3-d action can be shown to be strictly
equivalent to an effective pure scalar model. In this sense the model
provides motivating background for the investigation of the scalar
effective model in the case of a single scalar doublet (N=1).

\subsection{Effective Action in the Large N Limit}

Choosing the nonvanishing Higgs-field vacuum expectation value, $\Phi_0$,
in the first scalar doublet,
the computation naturally breaks up into a part identical with the calculation
of the reduced action for the model with one doublet and the other part which
consists of calculating the contribution from the appropriately separated
group of N-1 doublets. This statement is evident if  one considers the
quadratic part of the scalar action alone. The Higgs doublets are
parametrised
as:
\begin{equation}
\Phi_{q}=\left(\matrix{\Phi_{q1}\cr \Phi_{q2}\cr}\right)
          =\delta_{q,1}\left(\matrix{0\cr \Phi_0\cr}\right)+
         \left(\matrix{\xi_{q1}+i\xi_{q2}\cr
           \xi_{q3}+i\xi_{q4}\cr}
\right),~~~~~~q=1,...,N
\label{andras1}
\end{equation}
Their potential is O(4N)-symmetric and depends only on the length of the
4N-component Higgs-field vector:
\begin{equation}
U_{\rm Higgs}={1\over 2}m^{2}\Phi_q^{+}\Phi_q+{1\over 24}\lambda
        (\Phi_{q}^+\Phi_{q})^2.
\label{andras2}
\end{equation}
The quadratic action
derived with eq.(\ref{andras1}) and eq.(\ref{andras2}) has
the form:
\begin{equation}
S_{\rm Higgs}^{(2)}={1\over 2}(m^2+{\lambda\over 6}\Phi_0^2)
\sum_{q=2}^N\sum_{\alpha=1}^{4}\xi_{q\alpha}^{2}+{1\over 2}(m^2+
{\lambda\over 6}
\Phi_{0}^{2})\sum_{\alpha\neq 3}\xi_{1\alpha}^2+{1\over 2}(m^2+{\lambda\over 2}
\Phi_0^{2})\xi_{13}^2.
\end{equation}
Clearly, at 1-loop level the fluctuations of the fields with different
q are independent, and this is true also for the gauge-scalar coupling part
of the action.  In the quadratic part of the latter the gauge fields
couple only to the q=1 doublet, where the static part of the
Higgs-configuration was choosen to point to.  The fluctuations of the
other N-1 doublets coincide with those of 2(N-1) independent complex
scalar fields on a static abelian $A_0$ background, which is
characterised in the Fourier-space uniformly by the following $2\times
2$ matrix:
\begin{equation}
\left(\matrix{K^2+m^2+{g^2\over 4}A_0^2+{\lambda\over
6}\Phi_{0}^{2} &ig\omega_nA_0\cr -ig\omega_nA_0&K^2+m^2+{g^2\over
4}A_0^2 +{\lambda\over 6}\Phi_0^2\cr }\right),
\end{equation}
$(K^2={\bf k}^2+\omega_n^2)$. The
contribution from the coupled gauge-scalar fluctuations has been
evaluated before \cite{kajantie,jako}. We thus discuss here only the
contribution from the additional scalar fields:
\begin{equation}
\Delta U=(N-1)\sum_{n\neq 0}\int {d^3 k\over (2\pi)^3}
        \ln [(K^2+{g^2\over 4}A_0^2+m^2+
{\lambda\over 6} \Phi_0^2)^2-g^2\omega_n^{2}A_{0}^2].
\end{equation}
The usual expansion in the argument of the logarithm leads to modification of
the classical action, in which we are going to keep terms up to {\it dim 4}
combinations. After performing the frequency sum and the momentum integration
the following cut-off regularised corrections to the quartic
$\Phi_0 - A_0$-potential arise:
\begin{eqnarray}
&
\Delta U=(N-1)\{ {g^2\over 2}A_0^2(-{\Lambda T\over 2\pi^2}
+{T^2\over 6}+{m^2\over 8\pi^2})+{\lambda\over 3}\Phi_0^2({\Lambda^2\over
8\pi^2}-{\Lambda T\over 2\pi^2} \nonumber \\ & +{T^2 \over 12}+
{m^2\over 8\pi^2} - D_0{1\over 8\pi^2})
+{g^4\over 192\pi^2}A_0^4+{g^2\lambda\over 96\pi^2}A_0^2\Phi_0^2
+ {\lambda^2\over 36}\Phi_0^4({1\over 8\pi^2}-D_0{1\over 8\pi^2})\}
\label{andras3}
\end{eqnarray}
with $D_0=\ln{\Lambda\over T}$+const.
Adding eq.(\ref{andras3}) to the
1-doublet+gauge contribution, calculated in the
thermal static
gauge  previously \cite{jako}, the regularised full potential has the following
expression:
\begin{eqnarray}
&
U_{\rm static}[A_0,\Phi_0]={1\over 2}m^2\Phi_0^2+{1\over 24}\lambda\Phi_0^4
\nonumber \\ &
+{1\over 2}\Phi_0^2[({9\over 4}g^2+{2N+1\over 3}\lambda)
({\Lambda^2\over 8\pi^2}-{\Lambda T\over 2\pi^2}+{T^2\over 12}-
D_0{m^2\over 8\pi^2})
+{m^2\over 8\pi^2}({3g^2\over 4}+{(2N+1)\lambda\over 3})] \nonumber \\ &
+{1\over 24}\Phi_0^4({(N+2)\lambda^2\over 12\pi^2}+{27g^4\over
64\pi^2}+{3\lambda g^2\over 16\pi^2}-D_0({(N+2)\lambda^2\over 12\pi^2}+{27g^4
\over 64\pi^2}+{9\lambda g^2\over 16\pi^2})) \nonumber \\ &
+{N+16\over 192\pi^{2}}g^4A_0^4
+{1\over 2}A_0^2({N+4\over 6}g^2T^2+{Ng^2m^2\over 8\pi^2}-
{N+4\over 2\pi^2}g^2\Lambda T) \nonumber \\ &
+{1\over 8}g^2A_0^2\Phi_0^2(1+{(2N+1)\lambda\over 24\pi^2}+
{3g^2\over 16\pi^2}(5-3D_0)).
\end{eqnarray}
For the renormalised potential the "classical" extremum conditions are
prescribed at the classical position of the minimum \cite{linde}:
\begin{equation}
{dU(T-{\rm indep})\over d\Phi_0}=0,~~{d^{2}U({\rm T-indep})
\over d\Phi_0^{2}}=
m_{H}^{2}(T=0),~~~\Phi_0=v_{0}.
\label{num8}
\end{equation}
For the separation of the $T$-independent part one has to rewrite the
logarithmic part (proportional to $D_0$)
of the regularised potential conveniently:
\begin{eqnarray}
&
-{1\over 2}
[{1\over 16\pi^2}m^2\Phi_0^2
({9g^2\over 4}+{2N+1\over 3}\lambda)+{1\over 96\pi^2}\Phi_0^4
({(N+2)\lambda^2\over 3}+{27g^4\over 16}+{9\lambda g^2\over 4})]
\ln{\Lambda^2\over T^2} \nonumber \\ &
=-{1\over 64\pi^2}\sum_Q n_Qm_Q^4(\Phi_0)\ln{\Lambda^2\over T^2},
\label{num9}
\end{eqnarray}
Q=T(ransversal),L(ongitudinal),H(iggs),G(oldstone). The most natural choice for
$m_Q^2(\Phi_0)$ is the following:
\begin{eqnarray}
&&
n_T=6,~~~~~~m_T^2={1\over 4}g^2\Phi_0^2, \nonumber \\ &&
n_H=1,~~~~~~m_H^2=m^2+{\lambda\over 2}\Phi_0^2, \nonumber \\ &&
n_G=4N-1,~~~~~~m_G^2=m^2+{\lambda '\over 6}\Phi_0^2, \nonumber \\ &&
n_L=3,~~~~~~m_L^2= \biggl( {1\over 16}-{27\over 16(4N-1)}
\biggr)^{1/2}g^2\Phi_0^2
\end{eqnarray}
with $\lambda '=\lambda+27g^2/(8N-2)$. For large values of $N$ $m_L^2$
is real, and the limit $N \rightarrow \infty$ gives
\begin{equation}
n_L=3,~~~~~~m_L^2={1\over 4}g^2\Phi_0^2=m_T^2,~~~~(N=\infty)
\end{equation}
On the right hand side of eq.(\ref{num9}) one can then separate the cut-off
dependence from
the T-dependence by introducing appropriate normalisation scales,
possibly different for each value of Q. For the gauge coupling $g^2$ we
choose the renormalization scale $\mu =T$ and absorb the logarithmic and
finite corrections fully into its counterterm.
This is consistent with the omission of the effect of the wave-function
renormalisation in the present calculation (c.f. \cite{jako}).

After absorbing the T-independent infinities and some
finite parts into the counterterms of $m^2$, $\lambda$ and $g^2$ one finds the
following explicit T-dependent potential fulfilling the conditions eq.
(\ref{num8}):
\begin{eqnarray}
&
U[A_0,\Phi_0]={1\over 2}m^2\Phi_0^2+{1\over 24}\lambda\Phi_0^4
+{1\over 2}\Phi_0^2({9g^2\over 4}+{(2N+1)\lambda\over 3})({T^2\over 12}-
{\Lambda T\over 2\pi^2}) \nonumber \\ &
-{1\over 128\pi^2}m^2\Phi_0^2[{\lambda '(4N-1)\over 3}
\ln{9g^2v_0^2\over 16T^2}+ \lambda(4N-1)\ln{N\lambda v_0^2\over 3T^2}+
{27g^4\over \lambda}+{3g^2\lambda '\over 2\lambda}+2\lambda] \nonumber \\ &
+{1\over 1536\pi^2}\Phi_0^4[-{27g^4\over 2}\ln{Ng^2v_0^2\over 4T^2}-
{2(4N-1)\over 3}\lambda '^{2}\ln{9g^2v_0^2\over 16T^2}
\nonumber \\ & -6\lambda^2\ln{N\lambda
v_0^2\over 3T^2} +\lambda^2{8(N+2)\over 3}+{27\over 2}g^4+18g^2\lambda]
\nonumber \\ &
+{1\over 2}A_0^2({N+4\over 6}g^2T^2+{Ng^2m^2\over 8\pi^2}-
{N+4\over 2\pi^2}g^2\Lambda T) +{g^4\over 192\pi^2}A_0^4+{1\over 8}
g^2A_0^2\Phi_0^2
\label{num12}
\end{eqnarray}
($\langle \Phi^2 \rangle_{T=0}=Nv_0^2$).
The large N form of this potential depends crucially on the way the different
couplings scale with N:
\begin{equation}
g^2={\tilde g^2\over N^{\alpha}},~~~\lambda={\tilde\lambda\over N^{\beta}},
{}~~~m^2={\hat m^2\over N^\gamma} .
\end{equation}
By inspecting the second line of eq.(\ref{num12}) one realizes, following
the argument of \cite{linde}, that the stability of the symmetry
breaking vacuum at T=0 requires
\begin{equation}
{g^4\over \lambda}\leq{\cal O}(N^0)~~~{\rm or}~~~ \beta-2 \alpha\leq 0.
\end{equation}
This restriction excludes the proposition for the large N scaling law
suggested in
\cite{jain} $(\beta =1,\alpha =1/3)$. We continue our analysis along
the lines advocated in \cite{arnold}, which corresponds to
\begin{equation}
\alpha =1,~~~\beta =2,~~~\gamma =1.
\label{scaling1}
\end{equation}
The length of the $\Phi_0$-field is assumed
to be ${\cal O}(N^{1/2})$. We also note
that the choice $\alpha = 1$ for the scaling of
the gauge coupling agrees with similar prescriptions used for $SU(N_c)$
gauge theory coupled to $N_f$ fermions in the $N_f \rightarrow \infty$
limit \cite{banks}.

Substituting these scaling relations into eq.(\ref{num12}),
we obtain the following form of the leading large N
potential:
\begin{eqnarray}
&
U[A_0,\Phi_0]={1\over 2N}\Phi_0^2[\tilde m^2(1-{27\tilde g^4\over 64\tilde
\lambda\pi^2})+({9\tilde g^2\over 4}+{2\tilde\lambda\over 3})({T^{2}\over 12}-
{\Lambda T\over 2\pi^2})] \nonumber \\ &
+{1\over 24N^2}\Phi_0^4(\tilde \lambda +{27\tilde g^4\over 128\pi^2}
(1-\ln{\tilde g^2v_0^2\over 4T^2})) \nonumber \\ &
{1\over 2} A_0^2 ( {\tilde g^2\over 6} T^2 - {\tilde g^2\over 2\pi^2}
\Lambda T)+{1\over 8N}\tilde g^2A_0^2\Phi_0^2.
\end{eqnarray}
This result clearly
shows that under the scaling conditions of eq.(\ref{scaling1})
only
the result of the gauge integration survives the large N limit. This
conclusion agrees with the one obtained in Ref. \cite{arnold} concerning the
class of the leading Feynman graphs drawn in the full 4-d finite--T theory.
Each term of the potential is ${\cal O}(N^{0})$, that is the same order of
magnitude as that of the gauge part of the 3-d effective action. The latter one
becomes the copy of three effectively  Abelian gauge fields because $g^2$
now is scaling with an inverse power of N \cite{jain}. The
self-interaction of the 3-d isovector scalars $A_0$ belongs also to the
subleading
terms of the potential, therefore its action becomes formally quadratic, too.
With the (almost) usual 3-d-rescaling of the fields the leading N effective
action is given by the following expression
(the flavor index q has been omitted for convenience):
\begin{eqnarray}
&
{1\over 2}\int d^3x\{U_i^a[(-\partial_l^2+\tilde
g_3^2\phi^{+}\phi)\delta_{ij}
+(1-{1\over\alpha})\partial_i\partial_j]U_j^a+ 2i\tilde
g_3\sqrt{N}U_i^aJ_i^a \nonumber \\ &
U_0^a[-\partial_l^2+{\tilde g_3^2T\over 6}+{\tilde
g_3^2\phi^{+}\phi \over 4}]U_0^a+\phi^{+}[-N\partial_l^2+\hat
m^2]\phi +{\hat\lambda_3\over 12}(\phi^{+}\phi)^2\}\nonumber \\ &
+~~{\rm (3-d)-counterterms}
\end{eqnarray}
with
\begin{eqnarray}
&
J_i^a={1\over 4}(\partial_i\phi_q^{+}\tau^a\phi_q-\phi_q^{+}\tau^a
\partial_i\phi_q),\nonumber \\ &
\hat m^2=\tilde m^2(1-{27\tilde g_3^4\over 64\tilde \lambda_3\pi^2})+
({9\tilde g_3^2\over 4}+{2\tilde\lambda_3\over 3}){T\over
12},~~~\hat
\lambda_3 =\tilde \lambda_3+{27\tilde g^4\over 128\pi^2}(1-\ln{\tilde
g^2v_0^2\over 4T^2}),\nonumber \\ &
\tilde g_3^2=\tilde g^2T,~~\tilde\lambda_3=\tilde\lambda
T,~~U_i^a=A_i^a/\sqrt{T},~~
U_0^a=A_0^a/\sqrt{T},~~\phi=\sqrt{{1\over NT}}\Phi_0.
\end{eqnarray}
The integration over $U_i$ and $U_0$ can be performed trivially. For
instance in the Landau-gauge the non-regularised
contributions to the effective $\phi$-action reads:
\begin{equation}
3~{\rm Tr}~\log (-\partial_{l}^2+\tilde g_3^2\phi^{+}\phi )+
{3\over 2}{~\rm Tr}~\log(-\partial_{l}^{2}+{\tilde g_3^2 T\over 6}+
{\tilde g_3^2\phi^{+}\phi \over 4})+{N\over 2}\tilde
g_3^2J_i^aD_{ij}^{ab}J_j^b,
\end{equation}
where $D_{ij}^{ab}$ denotes the propagator of the transversal gauge
quanta on the scalar background. The last term is actually a correction
to the kinetic term of the Higgs-fields. Taking into account that the
fluctuations of $\phi$ are ${\cal O}(1/\sqrt{N})$, the magnitude of
this term is also unity.  After performing the 3-d functional traces
the 3-d mass-counterterms proportional to the gauge coupling
exactly cancel and the
following effective scalar theory is derived:
\begin{eqnarray}
&
S_{3d}[\phi ]=\int d^3x\{{1\over 2}\phi^{+}[-N\partial_l^2+\hat
m^2]\phi +{1\over 2}N\tilde g_3^2 J_i^aD_{ij}^{ab}J_j^b+{\hat\lambda_3\over 24}
(\phi^{+}\phi)^2 \nonumber \\ &
-{1\over 4\pi}(\tilde m_A^2+{\tilde g_3^2\phi^{+}\phi\over 4})^{3/2}-
{\tilde g_3^3\over 8\pi}({T\over 6}+{\phi^{+}\phi\over 4})^{3/2}
-{\tilde\lambda_3\Lambda\over 6\pi^2}\phi^{+}\phi\}.
\label{theory1}
\end{eqnarray}
In this equation we indicate, where could have appeared the magnetic
screening mass $\tilde m_A^2$, which to leading order in N is strictly
0.

The structure of the effective theory eq.(\ref{theory1}) agrees with the basic
structure of the scalar theory to be investigated in later parts of
this paper, up to the correction of the kinetic part which we shall
omit from our quantitative investigation.

\subsection{The Effective Model for $N$=1}

For $N$=1 one has to go back to the form of the potential term of the
3-d effective action given in eq.(\ref{num12}). This
form has been established in Ref.
\cite{jako}, where also
the 1-loop solution of the 3-d effective model has been extensively
discussed. The authors suggest the existence of an interesting
screening mass hierarchy at the temperature of the phase transition,
whose characterisation depends crucially on the value of $m_H(T=0)$.

Namely, the effective screening mass squares of the $A_0$ and $A_i$ quanta
given by the formulae
\begin{equation}
m_D^2={5\over 6}g^2T^2+{m^2g^2\over 8\pi^2}+{g^2\over
4}\Phi_0^2(T=T_c)
,~~~m_W^2={g^2\over
4}\Phi_0^2(T=T_c)
\label{num21}
\end{equation}
turn out to be  a factor 4-6 larger than that of the scalar quanta for
$m_H(T=0)$=35 GeV. This ratio monotonically increases as one goes to
smaller Higgs mass values, its value is $\sim 25$ for $m_H(T=0)$=20 GeV.
Therefore, at least in this region, the
integration over these very massive degrees of freedom
seems to be well-founded.

After gaussian integration and after
3-d mass-renormalisation of $A_0$ the following effective
action for the Higgs-field is obtained:
\begin{eqnarray}
&
S_{3d}[\phi ]=\int d^3x[{1\over 2}\partial_l\phi^{+}\partial_l\phi+{1\over
2}\hat M_{\phi}^{2}\phi^{+}\phi+{\hat\lambda T\over 24}(\phi^{+}\phi)^2
\nonumber \\ &
-{g_3^3\over 32\pi}({4m_D^2\over
g_3^2}+\phi^{+}\phi)^{3/2}-{g_3^3\over 16\pi}({4m_A^2\over
g_3^2}+\phi^{+}\phi)^{3/2}],
\label{num22}
\end{eqnarray}
with the meaning of the notations given below (not to be mixed with previous
meaning of some of them!):
\begin{equation}
\hat M_{\phi}^2=\hat m^{2}+({3g^2\over 16}+{\lambda\over
12})T^2-\lambda{T\Lambda\over 2\pi^2},
\label{num23}
\end{equation}
\begin{eqnarray}
&
\hat m^2=m^2\{1-{1\over 32\pi^2}[({9\over
2}g^2+\lambda)\ln{3g^2v_0^2\over 4T^2}+\lambda\ln{\lambda
v_0^2\over 3T^2}]-{1\over 128\pi^2} (45g^2+20\lambda+{27g^4\over \lambda})\},
\nonumber \\ &
\hat\lambda =\lambda- {9\over 16\pi^2}({9g^4\over
16}+{3g^2\lambda\over 4}+{\lambda^2\over 3}) \nonumber \\ &
-{3\over 8\pi^2}\{g^4({3\over 8}\ln{g^2v_0^2\over
4T^2}-{3\over 2}\ln{g^2v_0^2\over \sqrt{2}T^2})+{\lambda^2\over
4}\ln{\lambda v_0^2\over 3T^2}+3({3g^2\over 4}+{\lambda \over
6})^2\ln{3g^2v_0^2 \over 4T^2}\}.
\label{num24}
\end{eqnarray}

Eqs.(\ref{num22})-(\ref{num24}) represent the version of the effective scalar
model, which we are going
to employ in the non-perturbative investigations of the next sections.
Its general structure fully coincides with the result of the large N
calculation, providing some extra argument for its relevance.

For its discrete treatment it is most convenient to keep $g^2$ fixed and
to think of the effective action as being
defined by the couplings $\hat M_{\phi}$ and $\hat \lambda$. After the
performing the phase transition analysis one uses equations
eq.(\ref{num23}) and eq.(\ref{num24}) for finding {\it a posteriori} the true
couplings (or what is equivalent the Higgs mass).  The deviation of the
quantities with "hat" from "hatless" partners depends crucially on the
logarithmic terms appearing in eq.(\ref{num23}) and eq.(\ref{num24}). With
decreasing Higgs-mass they become increasingly important, making our
conclusions for this region sensitive to the renormalisation scheme,
implemented in the 1-loop calculations. For comparison we can consider
a choice of the renormalisation scale in eq.(\ref{num9}), where the
logarithmic terms are missing $(\mu = T)$. (The assumption of unchanged
tree-level relationship between the couplings and the W and  Higgs
masses is maintained somewhat arbitrarily.) The results for $T_c$ and
$\Phi_c\over T_c$ will be seen in the next section to agree with their
values found in the Linde scheme within 10-15\% for $m_H(T=0)$=35 GeV
when the mean field solution method is applied. The deviation sharply
increases for smaller values of the Higgs mass.

\section{Lattice Formulation and Mean Field Analysis of the
Effective Scalar Model}

 The discretisation of eq.(\ref{num22}) proceeds by introducing the
dimensionless field
\begin{equation}
\psi =\sqrt{a}\phi
\end{equation}
with $a$ being the lattice constant. Introducing also dimensionless masses
and temperature via the relations:
\begin{equation}
2m_{A}a= \sqrt{\gamma_{M}}g\Theta,~~~2m_{D}a=\sqrt{\gamma_{E}}g\Theta,~~~
\Theta =aT
\label{num26}
\end{equation}
one finds
\begin{eqnarray}
&
S_{3d,lat}=\sum_{x}\bigl[{1\over 2\kappa}\psi_{x}^{+}\psi_{x}+
{\hat\lambda\Theta\over 24}(\psi_{x}^{+}\psi_{x})^{2}-{1\over 2}\sum_{e}(
\psi_{x+e}^{+}\psi_{x}+\psi_{x}^{+}\psi_{x+e}) \nonumber \\ &
-{g^{3}\Theta^{3/2}\over 32\pi}(2(\gamma_{M}\Theta+\psi_{x}^{+}\psi_{x}
)^{3/2}+(\gamma_{E}\Theta +\psi_{x}^{+}\psi_{x})^{3/2})\bigr]
\label{latact}
\end{eqnarray}
with
\begin{equation}
{1\over 2\kappa}={1\over 2}\hat m^{2}a^{2}+{1\over 2}({3\over 16}g^{2}+
{\lambda\over 12})
\Theta^{2}-{C\over 2}\Theta\Sigma(L^{3})+3~~.
\end{equation}
Here $\Sigma (L^{3})$ denotes the lattice regularized version of the
linearly divergent integral $\int d^3p/(8\pi^3p^2)$,
giving rise to the linear divergence in eq.(\ref{num23}).  The perturbative
values of the constants $C$ and $\gamma_E$ are given by $C=\lambda$ and
$\gamma_E= 10/3$.
We emphasize once more that the input parameters into the lattice calculation
are $\kappa, \hat\lambda$ and g.

For each fixed value of $\Theta$ the discrete theory, eq.(\ref{latact}),
will pass through a phase transition at a certain value
$\kappa_{c}(\Theta )$ or
\begin{equation}
Z_c({\Theta}) = ({1\over 2\kappa_{c}(\Theta)}-3+{C\over 2}
\Sigma (L^{3})\Theta ){1\over \Theta^{2}}.
\label{num29}
\end{equation}
The continuum limit
is defined by the limit $\Theta\rightarrow 0$
of $Z_c(\Theta)$
providing a non-trivial value
$Z_0$.

When the value of $Z_0$ has been determined, eq.(29) is rewritten
as a relation between the unknown physical value of $T_c$ and $\lambda$:
\begin{equation}
{\hat m^2\over 2T_c^2}=Z_0-{1\over 2}({3\over 16}g^2+{\lambda\over 12}).
\label{num30}
\end{equation}
In eq.(\ref{num30}) $\hat m$ is replaced by the expression given in
eq.(\ref{num24}).
The other relation is provided by the definition of $\hat\lambda$ in
eq.(\ref{num24}).  From these two equations the physical values of the
Higgs mass and of the critical temperature can be determined.

In concrete terms, after introducing the dimensionless temperature
(in units of the T=0 vacuum expectation
value $v_0$) the following two equations are to be solved (for instance,
iteratively):
\begin{eqnarray}
&
\tau_c^2={\lambda\over 6}{1\over {3\over 16}g^2+{\lambda\over 12}-
2Z_0(\hat\lambda )}(1-f_1(\lambda ,\tau_c)), \nonumber \\ &
\lambda=\hat\lambda+f_2(\lambda ,\tau_c),
\label{num31}
\end{eqnarray}
with the functions
\begin{eqnarray}
&
f_1(\lambda ,\tau_c)={1\over 128\pi^2}(45g^2+20\lambda +{27g^4\over \lambda})
+{1\over 32\pi^2}[({9\over 2}g^2+\lambda)\ln{3g^2\over 4\tau_c^2}+\lambda
\ln{\lambda\over 3\tau_c^2}], \nonumber \\ &
f_2(\lambda ,\tau_c)={9\over 16\pi^2}({9g^4\over 16}+{3g^2\lambda\over 4}
+{\lambda^2\over 3}) \nonumber \\ &
+{3\over 8\pi^2}[g^4({3\over 8}\ln{g^2\over 4\tau_c^2}-
{3\over 2}\ln{g^2\over \sqrt{2}\tau_c^2})+{\lambda^2\over 4}\ln{\lambda\over 3
\tau_c^2}+3({3g^2\over 4}+{\lambda\over 6})^2\ln{3g^2\over 4\tau_c^2}]
{}.
\label{num32}
\end{eqnarray}
(Here the relation $m^2=-\lambda v_0^2/6$ has been used again.)

Having found $\lambda$ and $\tau_c$ the physical Higgs-mass and the
critical temperature are determined easily from the relations:
\begin{equation}
\tau_c^2={T_c^2\over v_0^2},~~~\lambda ={3m_H^2\over v_0^2}.
\label{num33}
\end{equation}

The simplest illustration of a non-trivial continuum limit $(Z\neq 0$) is
given by the mean field analysis of the system eq.(\ref{latact}).
The mean-field expression of the free energy density as a function of the
mean field order parameter $s$ is
\begin{equation}
F[s]=({1\over 2\kappa}-d)s^{2}+{\Theta \hat\lambda\over 24}s^{4}-{g^{3}\Theta
^{3}\over 32\pi}[2(\gamma_{M}+{s^{2}\over \Theta})^{3/2}+(\gamma_{E}
+{s^{2}\over\Theta})^{3/2}].
\end{equation}
The condition for the existence of a non-trivial $(s_{0}\neq 0)$ minimum is
written as
\begin{equation}
({1\over 2\kappa}-d){1\over \Theta^{2}}+{\hat\lambda\over 12}{s_{0}^{2}
\over \Theta}-{3g^{3}\over 32\pi}[(\gamma_{M}+{s_{0}^{2}\over\Theta})^{1/2}+
{1\over 2}(\gamma_{E}+{s_{0}^{2}\over \Theta})^{1/2}]=0.
\end{equation}
Introducing the scaled free energy $F[s]/\Theta^{3}$ and the scaled order
parameter $\bar s^{2}=s^{2}/\Theta$, one can eliminate the common term
$({1\over 2\kappa}-d){1\over \Theta^{2}}$
from the degeneracy condition for the free energy minima and determine
the position of the degenerate symmetry breaking minimum $\bar s_{0}$ from the
equation
\begin{eqnarray}
&
-{\hat\lambda\over 24}\bar s_{0}^{4}+{3g^{3}\bar s_{0}^{2}\over 32\pi}
[(\gamma_{M}+
\bar s_{0}^{2})^{1/2}+{1\over 2}(\gamma_{E}+\bar s_{0}^{2})^{1/2}]
\nonumber \\ &
-{g^{3}\over 32\pi}[2(\gamma_{M}+\bar s_{0}^{2})^{3/2}+(\gamma_{E}+
\bar s_{0}^{2})^{3/2}]+{g^{3}\over 32\pi}(2\gamma_{M}^{3/2}+\gamma_{E}^{3/2})
=0.
\end{eqnarray}
The quantity $Z_{c,mf}$ is calculated as
\begin{equation}
Z_{c,mf}=({1\over 2\kappa_c}-d)/\Theta^2=
-{\hat\lambda\over 12}\bar s_0^2+{3g^3\over 32\pi}[(\gamma_M+\bar
s_0^2)^{1/2}+{1\over 2}(\gamma_E+\bar s_0^2)^{1/2}].
\end{equation}
We note that the right hand side is independent of $\Theta$ and therefore
$Z_{0,mf}\equiv Z_{c,mf}$.
In Fig. 1) we show the mean field estimate of the quantity $Z$ for
$\gamma_B = 0,~1,~2$ (see eq.(\ref{ma}) for the meaning of $\gamma_B$) as a
function of $\hat m_H$. The rapid rise for small $\hat m_H$ leads uniformly
(independent of $\gamma_B$) to a "critical" $\hat \lambda$ at which the
right hand side of eq.(\ref{num30}) vanishes. In
the approximation $m^2=\hat m^2$ and
$\lambda = \hat \lambda$ this corresponds to a divergent critical
temperature. However, because of the uncertainties due to the
application of different renormalization conditions this phenomenon
might be an artefact of the above approximation. The data point
appearing in Fig. 1) is the result of our numerical simulation. The
analysis leading to it is described in section 4.

The order parameter discontinuity characterising the phase transition
can be expressed with help of $\bar s_{0}$  as
\begin{equation}
\Phi(T_{c})=\bar s_{0}T_{c}.
\end{equation}

A very contentful characterisation of the first order transition is given
by the surface tension, $\sigma$, between
coexisting ordered and disordered regions.
Exploiting the scaling behavior of the mean field solution one finds for it
in the thin wall approximation:
\begin{eqnarray}
&
\bar\sigma_{mf}={\sigma\over T_{c}^{3}}=\int_{0}^{\bar s_{0}}d\bar s
[2f(\bar s)]^{1/2},\nonumber \\ &
f(\bar s)={F(s)-F(0)\over \Theta^{3}}=Z_{0,mf}\bar s^{2}+{\lambda\over 24}
\bar s^{4}-{g^3\over 32\pi}[2(\gamma_{M}+\bar s^{2})^{3/2}+(\gamma_{E}+
\bar s^{2})^{3/2}]\nonumber \\ &
{}~~~~~~~~~~~~~~~+{g^{3}\over 32\pi}(2\gamma_{M}^{3/2}+\gamma_{E}^{3/2}).
\label{num39}
\end{eqnarray}

In Fig. 2)
we show the true Higgs mass square as a function of $\hat m_H^2=\hat
\lambda v_0^2/3$. The trivial linear relation goes over into a strongly
non-linear
functional form only for Higgs mass values less than 25 GeV. For larger values
$m_H$ and $\hat m_H$ are seen to agree within 10\%.
In Fig. 3)  $T_c$ is shown as a function of $m_H$ and $\hat m_H$.
Below $m_H=$25 GeV $T_c$ is rising with decreasing Higgs mass,
reflecting the approach to zero of the denominator in the expression of
$\tau_c$ in eq.(\ref{num31}).

The solid curve in this figure shows the result of the approximation
$m_H=
\hat m_H$, $\lambda =\hat\lambda$. This approximation corresponds to the
choice of normalisation scale $\mu =T$ in the interpretation of
eq.(\ref{num9}). It leads
to the definition of
the continuum limit directly from eq.(\ref{num30}) avoiding the
complicated procedure of solving eq.(\ref{num31}) and eq.(\ref{num32}).
This approximation for a Higgs mass near
$m_H$=15 GeV would indicate infinite $T_c$, which would hint to the
non-restorability of the broken gauge symmetry for Higgs masses smaller than
this limiting value. In principle, the non-restoration of the symmetry
even for infinite temperature is not excluded and there are known examples
of such behaviour due to quantum corrections to the effective potential
\cite{salomonson}.
These are the logarithmic terms of the expression of
$\hat \lambda$ and $\hat m^2$, which counterbalance the increasing tendency
of $T_c$, when the Linde-type renormalisation condition is being used.
However, large logarithmic corrections due to the essential deviation of the
relevant range of the temperature from the chosen renormalisation scale
make the 1-loop approximation unreliable. The
difference in the two curves of Fig. 3) for $m_H <$ 30 GeV indicates the
importance of higher loop contributions in the equations governing
the continuum limit.

In Fig. 4) the dependence of the order parameter discontinuities
on the strength of magnetic screening is shown as a function of $\hat m_H$.
For non-vanishing magnetic mass the first order transition ends in a
tricritical point (TCP). Following \cite{bucha} we parametrize the
amount of magnetic screening in proportion to the value obtained from
a selfconsistent Dyson-Schwinger equation:
\begin{equation}
m_A=\gamma_B{g^2T\over 3\pi}.
\label{ma}
\end{equation}
The mean field analysis yields
then for the end point values of the Higgs-masses:
\begin{eqnarray}
&&
\gamma_B=1~~,~~ m_H(TCP)=84.5~~ {\rm GeV},\nonumber \\&&
\gamma_B=2~~,~~ m_H(TCP)=61.7~~ {\rm GeV}
\end{eqnarray}
in good agreement with the continuum perturbative estimates. In the insertion
of Fig. 4) the uncertainty in the order parameter discontinuity resulting from
the use different renormalisation schemes is illustrated.
The application of the Linde-conditions leads for small Higgs masses
(dotted line) to harder transitions.
We can conclude, that in the region $m_H <$ 30-35 GeV
no reliable continuum statement can be made on the
basis of the 1-loop calculation.

Finally we display in Fig. 5) as function of $\hat m_H$ the mean field
result for the interface tension eq.(\ref{num39}).

\section{Numerical Simulation and Results}

   We have simulated the lattice model eq.(\ref{latact})
on hypercubic lattices with
linear extent $L$ and periodic boundary conditions.
The electric and magnetic screening mass parameters
were choosen
to be $\gamma_E=10/3$ and $\gamma_B=1$,  eq.(\ref{ma}).
At values for the W-mass $m_W=80$ GeV and for the zero
temperature vacuum expectation value of the Higgs field
$v_0=246$ GeV
the gauge coupling constant $g$ of eq.(\ref{latact})
is determined by the tree level relation
eq.(\ref{num21}) applied at $T=0$. These choices
first leave us with $3$ free parameters
of the theory $\hat\lambda,\Theta$ and the hopping parameter $\kappa$.
Choosing furthermore a value for the bare quartic coupling $\hat\lambda$
or correspondingly via
eqs.(\ref{num31},\ref{num32},\ref{num33})
a physical value for the Higgs mass $m_H$, we are
left in our simulation with a two parameter theory. In the $\kappa$-$\Theta$
plane
of couplings we expect the symmetry restoration phase transition
whose properties we would like to study. In detail
we shall discuss how to extract $T_c$, the order parameter discontinuity,
$\Phi(T_c)$, the latent heat $\cal L$ and the interface tension $\sigma$
from the raw data.

\subsection{Lattice Observables and the Continuum Limit}

Both bare parameters $\kappa$ and $\Theta$ serve the purpose of
constructing the continuum limit.
In particular we will be
interested in the study of various quantities measured at
$T_c$. We thus will tune the hopping
parameter to the critical line $\kappa_c(\Theta)$ and consider the
limit $\Theta \to 0$, which at the same time removes
the lattice cutoff $a$.
In this limit the theory turns into the $3$-dimensional Gaussian
model. From the $\Theta$-dependence of the critical
hopping parameter, we can determine the continuum limit of the quantity
$Z_c(\Theta)$, which has been defined in eq.(\ref{num29}). With the help of
eqs.(\ref{num31},\ref{num32},\ref{num33}) this
fixes the critical temperature in units of the Higgs
mass. We furthermore study various operators, which allow the extraction
of physically relevant quantities such as the order parameter
discontinuity at $T_c$, the surface tension and the latent heat. In
particular we consider the following observables
\begin{equation}
O_1={1\over L^3}\sum_x \psi_x^+\psi_x~~,~~
O_2={1\over 3L^3}\sum_{x,e} \psi_x^+\psi_{x+e}~~,~~
O_3={1\over L^3}\sqrt{\lbrace(\sum_{x} \psi_x)^+
                (\sum_{y} \psi_y) \rbrace},
\label{oper}
\end{equation}
whose distributions $P(O_i)$, expectation values $\langle O_i \rangle$
and diagonal fluctuations
$C(O_i)=L^3 \langle O_i^2-\langle O_i\rangle^2\rangle$ for
$i=1,2,3$ are determined in the simulation.
$O_1$ corresponds to the average length square of the Higgs field,
$O_2$ to a part of the kinetic term of the action, while
$O_3$ corresponds to the average field.  They can be related to the
corresponding continuum observables through relations discussed in the
previous sections. Here we note that the lattice fields $\psi$ and the
continuum fields $\Phi$ differ by a factor of $\sqrt{T}$. In the limit
$\Theta \to 0$ we thus expect appropriately scaled dimensionless ratios
to reach constant values. For instance we examine the ratio
${\langle O_3 \rangle \over \Theta^{1/2} } $, which in the limit $\Theta
\to 0$ converges to the continuum result $\Phi (T_c)/T_c$. Of course we
have to investigate here carefully possible scaling violations due to
finite lattice spacing $a$, as well as finite volume effects due to the
finite extent $L^3$ of the lattices. These problems will be discussed
in detail in connection with the numerical analysis of the various
observables.

In general we expect that with decreasing values of $\Theta$ (decreasing
lattice spacing) it will be increasingly difficult investigate the
properties of finite temperature phase transition as any signal for a
possible first order phase transition will become small in units of the
lattice spacing. Moreover, we have to understand the possible range of
values of the Higgs mass, for which we can study the phase structure of our
model. For large values of the Higgs mass the transition is expected to
be at best weakly first order, which then requires large lattices to
resolve any discontinuity in physical observables.
In order to explore the parameter range, in
which we can perform statistically significant studies of our model
we have studied thermal cycles across the phase transition line
$\kappa_c(\Theta)$ varying also the Higgs mass parameter.

 In Fig. 6) and Fig. 7) we display the results of thermal
cycles for the order parameter $\langle O_3 \rangle$
on  $18^3$ lattices with a statistics of $4000$ sweeps
per data point. Fig. 6) displays results at fixed
values of $\Theta=5$.
The parameter
$\hat m_H$ ranges from $\hat m_H=35$ GeV
to $\hat m_H=47$ GeV in this figure. The broad and pronounced hysteresis
at small values of $\hat m_H$ rapidly shrinks with increasing $\hat m_H$ and
disappears at $\hat m_H=47$ GeV. We have included into both figures
the mean field result for the critical point (dashed vertical lines)
and the mean field value for the order parameter discontinuity
(solid circles and height of the dashed lines). While
the locations of the phase transitions
roughly agree  with the centers of the hystereses, Fig. 6)
indicates a more rapid decrease of the order parameter jump
as compared to the mean field result.
At $\hat m_H=47$ GeV and
$\Theta=5$
it is presumably notoriously
difficult to determine infinite volume values of possible gaps
in operators, or to determine whether the phase tranition is
discontinuous at all. This is the main reason why we have
decided to choose the value
$\hat m_H=35$ GeV for a detailed and high precision study of
the symmetry restoration phase transition.
This value of $\hat m_H=35$ GeV
is still large enough for not being plagued too much in the
continuum interpretation by the ambiguities
of different renormalization schemes.
It corresponds to a physical value of the Higgs mass
of $m_H=37.16$ GeV, see Fig. 2).

Fig. 7) displays at the selected value of $\hat m_H=35$ GeV the variation
of the hysteresis of the order parameter $O_3$ as a function of $\Theta$.
As expected
the first order signal weakens with decreasing $\Theta$. Comparing
the mean field values of the order parameter jump
with the MC data
we again note a more
rapid weakening of the first order signal. From these data
and from additional MC-runs we expect that the interesting
region in which we might
be
able to resolve possible first order signals
on the given sized lattices extends down to $\Theta$-values
as low as $\Theta=1$.
The typical
lattices sizes, which we use in our simulation
range from $8^3$ up to $18^3$.
Typical statistics for simulations
on any of the lattice sizes is about $10^6$ sweeps.
To overcome the problem of large tunneling
times at the first order phase transition, we have used the
multicanonical ensemble approach \cite{multi}.
We see
later that even at $\Theta=3$ the phase
transition appears to be rather week, e.g. double
peak structures in the probability distributions
$P(O_i)$ at the transition point are not very pronounced.

\subsection{The critical Temperature $T_c$}

  The perturbative treatment of the effective
model predicts a certain shape of the phase transition
line $\kappa_c(\Theta)$, provided $Z_c(\Theta)$ of eq.(\ref{num29})
is known. However the mean field treatment
of the theory predicts that $Z_c(\Theta)\equiv Z_{0,mf}$ is
just a constant independent of $\Theta$. We therefore
might expect that in a first approximation
also the fully fluctuating theory
can be described with a
constant $Z_c(\Theta)=Z_{0,mc}$ for small values of $\Theta$.
At $\hat m_H=35$ GeV we estimate the critical hopping
parameter from thermal cycles. These are displayed
in Fig. 8) for $\Theta$ less than $1$.
We observe a clear bending of the phase transition line
towards smaller $\kappa_c$-values at $\Theta$-values of about unity.
Thus
the numerical data
predict a nonzero and positive value for $Z_{0,mc}$.
A fit to the shape of the phase transition line
according to eq.(\ref{num29}) then results into
a Monte Carlo determination of the quantity
$Z_{0,mc}=0.0187(16)$, which is to be compared with the mean field
value of $Z_{0,mf}=0.0113$ at same values of couplings.
Noticeably our data are consistent with the presence of
a linear $\Theta$-dependent term in eq.(\ref{num29}) with $C=0.93(16)
\times\lambda$
, as it is predicted by perturbation theory (the continuous curve in
the figure).  The mean field phase transition line (dashed curve)
deviates from the MC data, which at small $\Theta$ slightly overshoot
the $\kappa$-value $\kappa={1\over 6}$.  At the given value of
$Z_{0,mc}$ we may now determine the critical temperature for both of
our renormalization schemes, namely in the $\mu =T$ scheme (without
logarithmic corrections), as well as in the Linde-scheme, which
incorporates them.  The result is
\begin{eqnarray}
&&
T_c=114.9(36)~GeV~~~(\mu =T) \nonumber \\  &&
T_c=114.3(30)~GeV~~~{\rm Linde~scheme}.
\end{eqnarray}
Both numbers are displayed in Fig. 3), the triangle corresponds
to the $\mu =T$-scheme and is plotted at its value of $\hat m_H$
while the circle corresponds to the renormalization
scheme with logarithmic corrections plotted at the the physical Higgs
mass value $m_H$. It is noticeable that the difference between the schemes is
very small.
These Monte Carlo determinations of $T_c$ should be
compared with
the mean field result $T_c=99.6~GeV$.

\subsection{Critical Hopping Parameters}

For the value of $\hat m_H=35$ GeV, we have performed
a detailed finite size scaling analysis
of the symmetry restoration phase transition at three values of $\Theta$,
namely $\Theta=3,4$ and $5$.
In Fig. 9) we compare the results of a thermal cycle on an $18^3$
lattice for the operator $\langle O_1 \rangle$, with results of multicanonical
simulations for the same operator on $8^3,10^3$ and $12^3$ lattices.
The mean field position of the critical point, as well as
its true infinite volume value, deduced from finite size
scaling analysis, are indicated by the vertical dashed lines in the figure.
Defining pseudocritical points $\kappa_{max}(L)$ from the peak
positions of the maxima of the diagonal fluctuation $C(O_1)$, see Fig. 10),
the pseudocritical $\kappa$-values are extrapolated to infinite
volume via
\begin{equation}
\kappa_{max}(L)=\kappa_c + {b \over L^3}.
\end{equation}
Fig. 11) shows at $\Theta=4$ the finite size
scaling analysis leading to the infinite volume critical point.
It is clear from these figures that multicanonical
ensemble simulations combined with finite size scaling
theory allow a precise determination
of the location of the phase transition
or discontinuities. They are in fact superior
to standard methods e.g., thermal cycles, which even on a large $18^3$
lattice cannot produce numbers with comparable precision, see
Fig. 9).
Table 1) then contains for the selected $\Theta$-values
a comparison of the mean field critical hopping parameters
to the Monte Carlo results. They are very close and their
difference is  ${\cal O} (10^{-4})$.
As one might have expected the location
of the phase transition at large values of $\Theta$ is well described
by mean field behavior. This has to be contrasted with the observed behavior
of $\kappa_c(\Theta)$ at values of $\Theta<1$, where the mean field
approximation is not adequate.
\begin{table}[h]
\begin{center}
\caption[tab1]{The critical hopping parameters $\kappa_c(\Theta)$,
to the left: mean field results, to the right: MC results.}
\vspace{2ex}
\begin{tabular}{||c|c|c||}                        \hline
 $\Theta$ & $\kappa_c(\Theta)_{mf}$ & $\kappa_c(\Theta)_{mc}$  \\ \hline
 3        & 0.161188     &  0.161112(4)  \\ \hline
 4        & 0.157170     &  0.157065(5)  \\ \hline
 5        & 0.152283     &  0.152170(2)  \\ \hline
\end{tabular}
\end{center}
\end{table}

\subsection{Determination of Discontinuities}

We obtain the infinite volume discontinuities
in thermodynamic quantities by analyzing the volume dependence
of probability distribution
functions $P(O_i)$ of operators $O_i$ defined in eq.({\ref{oper}).
In the vicinity of the transition
they develop double peak structures.  We
determine the values of the couplings, where in the finite volume system
the two peaks in $P(O_i)$ are of equal height.  This defines e.g., at fixed
value of $\Theta$, a value of the hopping parameter $\kappa_{eh}(L)$.
Fits to the maxima of the distribution functions then yield
finite volume estimators for the values of operators
in the metastable states, denoted by $\langle O_i\rangle_{SB}(L)$
(symmetry broken) and $\langle O_i\rangle_S(L)$ (symmetric) for $i=1,2,3$.  The
discontinuities
$\Delta\langle O_i\rangle (L)=\langle
O_i\rangle_{SB}(L)-\langle O_i\rangle_S(L)$ are then extrapolated to
infinite volume using standard
finite size scaling arguments \cite{first}.  Their
large volume $L$-dependence is expected to scale with the volume,
\begin{equation}
\Delta\langle O_i\rangle (L)=\Delta\langle O_i\rangle + {a \over
L^3}~~~i=1,2,3.
\label{fsize}
\end{equation}
In Fig. 12) we display on $10^3$ lattices probability
distributions of the operator $O_1$ i.e., the average length square of the
scalar field at $\kappa_{eh}(L)$ for
$\Theta$ values $\Theta=3,4$ and $5$. As can be seen
one finds clear double peak distribution functions, which can
actually be observed for all of our operators $O_i$ $i=1,2,3$.
Remarkably, however, the value of the minimum of the distribution functions
rapidly rises with decreasing $\Theta$. At $\Theta=3$ the
suppression of mixed phase states is hardly noticeable, which makes
the phase transition a rather weak first order phase transition already.

Fig. 13) shows the finite size scaling analysis of $\langle O_3\rangle_{SB}(L)$
corresponding to the order parameter jump in the symmetry broken
phase.
A fit to the data
with the form eq.(\ref{fsize}) readily gives
the infinite volume discontinuity $\Delta\langle O_3\rangle $.
Fig. 14) shows the same kind of
analysis for the operator $O_1$, corresponding to the
average length square of the scalar field.
In this case $\langle O_1\rangle_S(L)$ also has to be considered
for the calculation of the corresponding discontinuity.
Carrying out
the analysis for all our operators we arrive in Table 2)
at the infinite volume values for the discontinuities
for all considered operators and $\Theta$-values.
They can be compared with the mean field order parameter jump which
appears also in Table 2).
In Table 2) we present
the discontinuities in $O_1$ and $O_2$ scaled
by the square of the discontinuity in $O_3$, the order parameter.
They are consistent with a value of $1$. Thus actually
the various discontinuities are not independent at the transition, but
trivially connected. This is exactly what one expects from a mean field
type of behavior at the transition.

The figures also exhibit some
unusual finite size scaling behavior e.g., it can be noted that with
increasing lattice size the finite volume estimators of the
discontinuities increase. Many of the first order phase transitions in
statistical physics and lattice gauge theories exhibit an opposite
behavior, that is with increasing lattice size
the values of the obtained discontinuities decrease.
Figs. 13) and 14) also indicate that finite volume corrections become
rather large in size at the smaller values of $\Theta$ (see the
different slopes in the same figures). This is especially true for the
symmetry broken state and we attribute this property to the vicinity of
the Gaussian fixed point.

\begin{table}[h]
\begin{center}
\caption[tab1]{Values of the discontinuities.}
\vspace{2ex}
\begin{tabular}{||c|c|c|c|c||} \hline
 $\Theta$ & $s_{mf}$ & $\Delta
\langle O_3\rangle $ & ${\Delta\langle O_1\rangle
\over \Delta\langle O_3\rangle^2}$
                                      &
${\Delta\langle O_2\rangle \over \Delta\langle O_3\rangle^2}$ \\ \hline
 1        & 1.312    & -              & -         & -         \\ \hline
 3        & 2.285    & 1.603(50)      & 0.955(70) & 0.946(69) \\ \hline
 4        & 2.639    & 2.036(05)      & 0.998(07) & 0.999(07) \\ \hline
 5        & 2.951    & 2.458(05)      & 0.991(06) & 1.000(07) \\ \hline
\end{tabular}
\end{center}
\end{table}

\subsection{The Order Parameter Jump}

  In Fig 15) we show as a function of $\Theta$
the order parameter jump $\Delta \langle O_3 \rangle$ scaled by the
mean field result. The actual magnitude of the order
parameter jump comes out smaller, than predicted by the mean field analysis.
In addition we also observe a mild deviation from scaling and
it appears as if possible scaling deviations
might be parametrized
by a correction linear in $\Theta$. Fig. 15) contains
a corresponding fit (dotted line in the figure), which predicts
that the order parameter jump in units of the mean field
order parameter discontinuity
might
in the continuum limit ${\Theta \to 0}$
be as low as one half.
However, the theoretical status of scaling
corrections in $3$-dimensions is unclear \cite{Zinn} and in our
context we might regard such an extrapolation as a lower bound
to the true value of the order parameter jump. Further
simulations closer to the
continuum are therefore required.
They may show a bending of the order parameter discontinuity to somewhat
larger values than half of the mean field result, but will be significantly
lower than the mean field prediction.
Such a result is consistent with the variational upper bound nature of
the mean field approximation.
Employing our extrapolation we currently estimate
the order parameter discontinuity to be
\begin{equation}
{\Phi(T_c) \over T_c}=0.68(4).
\end{equation}
Using the $T_c$ determination we obtain $\Phi(T_c)=78(5)$ GeV
at a physical Higgs mass value of $m_H=37.16~GeV$.

\subsection{The Latent Heat}

The volume normalized latent heat in units of $T_c^4$,
${\cal L} \over T_c^4$, is obtained from
the expectation value of the internal energy ${\cal U}\over T^4$
\begin{equation}
\langle {{\cal U}
\over T^4}\rangle =\langle
{1 \over \Theta^3} \beta \partial_\beta S(\beta)\rangle
\end{equation}
by calculating its difference in the symmetric
and the symmetry broken state at $T_c$.
\begin{equation}
{{\cal L}\over T_c^4} = \langle {{\cal U} \over T_c^4}\rangle \mid_{S}
-\langle {{\cal U} \over T_c^4}\rangle \mid_{SB}
\end{equation}
Hereby  $\beta$ denotes the inverse temperature $\beta={1\over T}$
and $S$ denotes the action of the $3+1$ dimensional theory.
We note that our convention in defining the latent heat is opposite
to what we have used for the order parameter discontinuities.
This makes the latent heat positive.
Incorporating the full temperature dependence
of all of our physical couplings we obtain for the
internal energy in terms of the $3$-dimensional reduced theory
the expression
\begin{eqnarray}
&
\langle {{\cal U} \over T^4}\rangle =  \langle
{S_{3D,lat}\over \Theta^3}
+\sum_x \lbrace
-( {3g^2\over 8}+{\lambda\over 6}-{m_H^2\over 16\pi^2 T^2}
({9\over 4}g^2+\lambda )  {\psi_x^+\psi_x \over \Theta} \nonumber \\ &
- {1\over 32\pi^2} ({9g^4 \over 16}+{\lambda^2 \over 3}+
{3 \lambda g^2 \over 4})
({\psi_x^+\psi_x \over \Theta})^2 \nonumber \\ &
+{g^3 \over 16\pi} [ ( ({2\gamma_B g \over 3\pi})^2+{\psi_x^+\psi_x
\over \Theta})^{3 \over 2}
+{1 \over 2}({10 \over 3}+ {\psi_x^+\psi_x \over \Theta} )^{3 \over 2} ]
\nonumber \\ &
+ {3g^3 \over 8\pi} [ ({\gamma_B g\over 3\pi})^2
(({2 \gamma_B g \over 3\pi})^2+{\psi_x^+\psi_x \over \Theta})^{1 \over 2}
+ {5 \over 12} ({10 \over 3}+{\psi_x^+\psi_x\over \Theta})^{1 \over 2}
]
\rbrace \rangle
\end{eqnarray}
We note here that the latent heat calculation
requires the calculation of discontinuities of more complicated operators
e.g., the calculation of discontinuities in
$\sum_x(\psi_x^+\psi_x)^2$ and discontinuities in certain
functions of the fields.
Since we did not measure all required
operators directly, we assume in the numerical analysis that
corresponding discontinuities show a mean field like behavior e.g., the
discontinuity in the operator $\sum_x(\psi_x^+\psi_x)^2$, for instance,
is replaced by the discontinuity in the operator $O_3$ to the fourth
power.  Our data support such an assumption, at least, for the restricted
set of operators at the large values of $\Theta$, considered in the
present investigation, see Table 2) and the related discussion.

In Fig. 16) we display the discontinuity of the volume normalized
$3$-dimensional
action eq.(\ref{latact}) scaled by corresponding $\Theta$-powers
at the critical point (circles) as a function of $\Theta$. It is
consistent with being zero and results from a cancellation of positive
and negative terms, which can be attributed to the kinetic term
and the potential term in the action. As the action gap
itself is related to the pressure, which is continous even
at the first order phase transition, this is an expected behavior
and merely is a check on the consistency of our analysis.

The latent heat per unit volume ${\cal L}\over T_c^4$
also appears in Fig. 16). The triangles denote the results
of our Monte Carlo simulation, while
the crosses (constant with $\Theta$) correspond
to the mean field analysis of the theory. Again we
observe a significant weakening of the phase transition
as compared to the mean field treatment.
We also observe scaling deviations i.e., as a function
of $\Theta$ the latent heat values have not yet
settled to their continuum value.
Here we quote the value of the latent heat obtained
from simulations at $\Theta=3$:
\begin{equation}
{{\cal L}\over T_c^4}=0.122(8).
\end{equation}
This value again is less than half of the mean field result
${{\cal L}\over T_c^4}=0.262$ and should be close to
the continuum result provided further scaling deviations turn out
not to be too large.

\subsection{The Interface Tension}

Following Binder \cite{Binder} the interface tension
inbetween symmetric and symmetry broken states of the theory
can be obtained from the shape of the
probability distribution functions at $\kappa_{eh}$.
Denoting the value of the maxima of the probability
distribution functions $P^{max}(O_i)$, states inbetween
the two maxima correspond
to mixed phases, which on the largest lattices form an interface with a
cross sectional
area $L^2$. Because of the periodic boundary conditions
there are actually two interfaces in the system and the total
interfacial area thus is $2L^2$.
Denoting the value of the probability distribution function
at its minimum $P^{min}(O_i)$, the interface tension can be defined
by the limit
\begin{equation}
\sigma_{lat} = \lim_{L \to \infty}  {1\over 2L^2}
{{\rm ln}[{P^{max}(O_i)\over P^{min}(O_i)}]}~~~i=1,2,3.
\label{binder}
\end{equation}
The quantity $\sigma_{lat}$ is connected to the continuum
interface tension $\sigma$ via $\sigma_{lat}={\sigma\over T_c^3}\Theta^2$.
It is however expected that the extrapolation to infinite
volume is more subtle, than in the case of discontinuities in the
order parameter.
Interfaces in finite boxes show sizable fluctuations controlled by their
stiffness and finite volume corrections to $\sigma_{lat}$ may have
a complicated $L$-dependent analytical form, as is demonstrated
by numerical simulations in spin models \cite{ourface}.
In Fig. 17) and Fig. 18) we exhibit our
interface tension analysis. Fig. 17) displays
the quantity
${{\rm ln}({P^{max}(O_1)\over P^{min}(O_1)})}$ for three
$\Theta$-values
as a function of the expected interfacial
cross section $L^2$ for all considered lattice sizes.
In particular for $\Theta=5$ we observe a very fast
rise with $L^2$, indicating that
mixed phase configurations are suppressed by many order of magnitudes
in the path integral and that the interface tension thus is large.
The data are also
roughly consistent with a linear increase of the considered quantity
with the interfacial cross section, though some curvature to larger
slopes is indicated. Since the status of possible finite volume correction
terms to our measured quantities is unclear, we may regard the straight
line fits of Fig. 17) as estimates for the expected surface tensions
values.  They are displayed in Fig. 18) as functions of $\Theta$ in
units of the corresponding mean field result obtained in the thin wall
approximation, eq.(\ref{num39}). They
show indeed a dramatic difference. In the mean field
calculation we expect at the given couplings a behavior according to
$\sigma_{lat}=0.0242~\Theta^2$.  Definitely large scaling deviations
are observed, while at the same time e.g., at $\Theta=3$, our estimates
are about $37$ times smaller than the corresponding mean field
calculation result.  Thus it is indicated here, that the interface
tension is subject to large corrections due to the
fluctuations of the scalar field.  Turning our interface calculation into
physical numbers the data point at $\Theta=3$ predicts an interface
tension of as low as about $960~{\rm GeV}^3$, which in the current context
may be viewed as an order of magnitude estimate.  The numerical
situation is not totally satisfying and in the future one should also
explore other possibilities for interface tension determinations, like
simulations on asymmetric lattices and studies of
tunneling masses on such geometries.

\hfill\break

\section{Conclusion}

In this paper a pure scalar model has been proposed for the description
of the finite temperature electroweak phase transition (EWPT). This
effective model was obtained from the full 3+1 dimensional field theory
through integration over the non-static Matsubara modes, realised in
the 1-loop approximation. The remaining 3-dimensional gauge degrees of
freedom and the effective adjoint Higgs-field were eliminated in a
subsequent Gaussian integration step.

Intuitive justification  for such scenario is provided by the thermal
mass hierarchy observed in the perturbative treatment of EWPT for
moderately small Higgs masses. Also, we have demonstrated that in an
appropriately defined large N limit the second integration step is exact.

The resulting model is a 3-dimensional O(4)-invariant  ferromagnet with
cubic and quartic potential.

Its discretisation and the proper procedure for taking the continuum limit
of its phase  transition have been thoroughly explained. Especially,
the sensitivity of
the $T_c$ dependence on $m_H(T=0)$ stemming from the application
of different renormalisation conditions in the continuum
has been discussed in detail.

The mean field solution of the lattice system and its continuum limit
has reproduced in all aspects the results of the improved 1-loop
perturbative treatments of EWPT [2,3] (influence of magnetic screening,
order parameter discontinuity, surface tension, etc.).  Also
it called attention to the dramatic sensitivity of $T_c$ to
the renormalisation conditions for $m_H(T=0)< 30$ GeV. This circumstance
has prevented the discussion of the possible existence of a lower
limiting Higgs mass value below which the gauge symmetry could not be
restored at any finite temperature.

The mean field analysis
served for orienting the numerical calculations, performed at
$m_H(T=0)\sim 35$ GeV. This value has been selected : i) to have a
strong signal of a first order transition, ii) to be free
of the uncertainties introduced by the renormalisation prescriptions ,
iii) to be able to compare the results of a careful numerical study
with other computer studies of EWPT based on different effective 3-d models.
\cite{kajantie}

The numerical data consistently exhibit at the considered
value of the Higgs mass a
weaker first order phase transition
as compared to the zeroth order mean field approximation
of the theory. Discontinuities of the theory like
the order parameter jump and the latent heat
turn out to be somewhat smaller than corresponding mean field
results. The critical temperature $T_c=114$ GeV appears
larger than mean field, again consistent with a weakening
of the first order phase transition. The interface
tension is strongly affected by the fluctuations
of the scalar field and its value will be much lower
than corresponding mean field predictions. By themselves
these findings may not be regarded as surprises as naively
one expects the fluctuations of the Higgs field to work
in the observed direction.
Similar experience has been gained from numerical estimations of the
interface tension of the SU(3) pure gluon theory.

Comparing our results
with the results of another MC simulation making use of another variant
of
$3$-dimensional effective models of the electroweak phase transition
at comparable value of the Higgs mass, we make, however, an interesting
observation. While the authors of \cite{kajantie} keep the gauge fields
in the $3$-dimensional Gauge-Higgs model as dynamical degrees
of freedom in their simulation, their findings indicate
a strengthening of the first order phase transition,
when compared to a one-loop perturbative analysis of their model,
and also, when compared to the results of our simulation.
They do quote e.g., a lower value of $T_c=85$ GeV at a value
of $m_H=35$ GeV or a value of the
order parameter discontinuity of about $170$ GeV, while
our value
$\Phi(T_c)=78(5)$ GeV is about half as large.
Thus it appears as if the non-perturbative dynamics
of gauge fields
in the dimensionally reduced model works in a direction opposite
to the effect of the fluctuations of the Higgs field, and eventually
dominates the physical characteristics of the phase transition.

\bigskip

{\bf Acknowledgments:} We thank the computer centers
at HLRZ, RWTH-Aachen and at the University of Cologne
for providing us with computer time on their
vector machines.  We are indebted to W. Buchm\"uller
for a useful discussion. We also acknowledge the support of the
EEC through contract numbers ERB-CHRX-CT-92-0051
and ERB-CIPA-CT-92-2061.
\hfill\break

\vfill\eject

\end{document}